\documentclass[preprint,prb]{revtex4}
\usepackage{graphicx}
\usepackage{latexsym}
\usepackage{amsfonts}
\begin{document}

\title{Effective \textit{gluon} dynamics in a stochastic vacuum}

\author{Jose A. Magpantay}
\email{jamag@nip.upd.edu.ph}

\affiliation{National Institute of Physics, University of the
Philippines, Diliman Quezon City, 1101,Philippines}

\date{\today}

\begin{abstract}
Using the new scalar and vector degrees of freedom derived from
the non-linear gauge condition $(\partial\cdot D)(\partial\cdot
A)=0$, we show that the effective classical dynamics of the vector
fields (identified as ``gluons'') in the stochastic vacuum defined
by the scalars result in the vector fields acquiring a range of
possible masses and losing their self-interactions.  From this
range of masses, we derive the mass gap in pure Yang-Mills theory.
Finally, we comment on the gauge-invariance of the result.
\end{abstract}

\maketitle

\section{Introduction}

The non-linear gauge condition
\begin{equation}\label{1}
(\partial\cdot D)^{ab}(\partial\cdot A^{b})=(D\cdot\partial)^{ab}(\partial\cdot A^{b})=\frac{1}{2}[(D\cdot\partial)+(\partial\cdot D)]^{ab}(\partial\cdot A^{b})=0,
\end{equation}
was proposed by the author\cite{gauge} as a generalization to the Coulomb gauge because the non-linear sector of (1), i.e., those configurations that satisfy $\partial\cdot A^{a}=f^{a}(x)\neq 0$, cannot be gauge transformed to the Coulomb gauge\cite{grav}. Subsequent papers established that:\\

(a) The gauge condition interpolates between the Coulomb gauge (with transverse fields in the short-distance regime) and the quadratic condition (with the new scalar and vector fields) in the large-distance regime\cite{low}.\\

(b) The gauge condition can no longer be extended to that with higher orders of the Fadeev-Popov operator, i.e., $(\partial\cdot D)^{n}(\partial\cdot A)=0$ for $n=2$,... is not consistent\cite{low}.\\

(c) The effective action of the scalar degrees of freedom is infinitely non-linear and a stochastic treatment of a class of classical configurations, those that are spherically symmetric in 4D Euclidean space-time, leads to the linear potential and the area law behaviour of the Wilson loop\cite{mass}.\\

(d) The full quantum treatment of the scalar degree of freedom shows that its dynamics is equivalent to that of a 2D O(1,3) non-linear sigma model\cite{free}.\\

In this paper, the effective classical dynamics of the vector field, which we will identify as the ``gluons'', in the stochastic background defined by the spherically symmetric scalar will be considered.  We will show that the ``gluons'' will acquire a mass and become non self-interacting.\\

This paper is then arranged as follows:  In Section II, we give the basic equations that govern the new degrees of freedom derived from the non-linear gauge.  Section III focuses on the effective dynamics of the ``gluons'' and derives an expression for the masses and the mass gap in pure Yang-Mills theory.  Section IV introduces the notion of limited gauge-invariance and justifies the gauge-invariance of the results derived.  Section V concludes with a comparison of the ideas presented here and those found in the literature.  The details of some of the computations are given in the Appendices

\section{The Dynamics of the New Degrees of Freedom}

Consider $SU(2)$ Yang-Mills in 4D Euclidean space-time.  Field configurations that satisfy equation (1) with $\partial\cdot A^{a}=\frac{1}{g\ell^{2}}f^{a}$ can be decomposed as:
\begin{equation}\label{2}
A^{a}_{\mu}(x)=\frac{1}{(1+\vec{f}\cdot\vec{f})}(\delta^{ab}+\epsilon^{abc}f^{c}+f^{a}f^{b})(\frac{1}{g}\partial_{\mu}f^{b}+t^{b}_{\mu}),
\end{equation}
where $f^{a}$ and $t^{a}_{\mu}$ are the new scalar (dimensionless) and vector (dimension 1) degrees of freedom.  These new degrees of freedom satisfy
\begin{eqnarray}\label{3}
\partial_{\mu}t^{a}_{\mu}&=&\frac{1}{g\ell^{2}}f^{a},\\
\rho^{a}&=&\frac{1}{(1+\vec{f}\cdot\vec{f})^{2}}[\epsilon^{abc}+\epsilon^{abd}f^{d}f^{c}-\epsilon^{acd}f^{d}f^{b}+f^{a}f^{d}\epsilon^{dbc}\nonumber\\
&-&f^{a}(1+\vec{f}\cdot\vec{f})\delta^{bc}-f^{c}(1+\vec{f}\cdot\vec{f})\delta^{ab}]\partial_{\mu}f^{b}t^{c}_{\mu}=0,
\end{eqnarray}
which make the number of degrees of freedom tally with those of $A^{a}_{\mu}$.\\

Note that the non-linear gauge has a built-in length scale $\ell$.  In the previous papers, $\ell$ was simply introduced for dimensional reason, i.e., since $\partial\cdot A^{a}\sim (length)^{-2}$, $\ell$ is needed to make $f^{a}$ dimensionless.  Naively, since the entire point of the non-linear regime of (1) is confinement, then we should identify $\ell$ to be of the order of 1 fermi.  But is there a stronger basis for $\ell$?\\

The non-linear gauge condition (see equation (1)) can be written as
\begin{equation}\label{4}
(\Box^{2}\delta^{ab}-g \epsilon^{abc}A^{c}\cdot\partial)(\partial\cdot A^{b})=0.
\end{equation}
As we will argue in Section V, as we let the coupling constant $g$ runs as given by
\begin{equation}\label{5}
\alpha = \frac{g^{2}(\mu)}{4\pi}\quad \sim \quad \frac{1}{\ell n(\frac{\mu}{\Lambda_{QCD}})}
\end{equation}
and $A^{a}_{\mu}$ scans the configuration space, we get to the regime where $(\partial\cdot D)^{ab}$ is singular with $\partial\cdot A^{a}$ itself as the zero mode.  This happens when $g(\mu)$ is strong enough so that $-g(\mu)\epsilon^{abc}A^{c}\cdot\partial$ is able to lower one of the positive eigenvalues of $\Box^{2}$ to zero eigenvalue.  Thus, confinement occurs at $g(\mu\sim\Lambda_{QCD})$, and we identify the length scale $\ell$ equal to $\frac{1}{\Lambda_{QCD}}$\\

Substituting equation (2) in $F^{a}_{\mu\nu}$, we find that
\begin{equation}\label{6}
F^{a}_{\mu\nu}=\frac{1}{g}Z^{a}_{\mu\nu}(f)+L^{a}_{\mu\nu}(f;t)+g Q^{a}_{\mu\nu}(f;t),
\end{equation}
where Z, L and Q are all functions of f and zeroth order, linear and quadratic in $t^{a}_{\mu}$, respectively.  Explicitly, the functions are:
\begin{eqnarray}\label{7}
Z^{a}_{\mu\nu}&=&X^{abc}\partial_{\mu}f^{b}\partial_{\nu}f^{c},\\
L^{a}_{\mu\nu}&=&R^{ab}(f)(\partial_{\mu}t^{b}_{\nu}-\partial_{\nu}t^{b}_{\mu})+Y^{abc}(\partial_{\mu}f^{b}t^{c}_{\nu}-\partial_{\nu}f^{b}t^{c}_{\mu}),\\
Q^{a}_{\mu\nu}&=&T^{abc}(f)t^{b}_{\mu}t^{c}_{\nu},
\end{eqnarray}
and the coefficient functions are
\begin{eqnarray}\label{8}
X^{abc}&=&\frac{1}{(1+\vec{f}\cdot\vec{f})^{2}}[-(1+2\vec{f}\cdot\vec{f})\epsilon^{abc}+2\delta^{ab}f^{c}-2\delta^{ac}f^{b}\nonumber\\
&+&3\epsilon^{abd}f^{d}f^{c}-3\epsilon^{acd}f^{d}f^{b}+\epsilon^{bcd}f^{a}f^{d}],\\
R^{ab}(f)&=&\frac{1}{(1+\vec{f}\cdot\vec{f})}(\delta^{ab}+\epsilon^{abc}f^{c}+f^{a}f^{b}),\\
Y^{abc}&=&\frac{1}{(1+\vec{f}\cdot\vec{f})^{2}}[-(\vec{f}\cdot\vec{f})\epsilon^{abc}+(1+\vec{f}\cdot\vec{f})f^{a}\delta^{bc}-(1-\vec{f}\cdot\vec{f})\delta^{ac}f^{b}\nonumber\\
&+&3\epsilon^{cad}f^{d}f^{b}-2f^{a}f^{b}f^{c}+\epsilon^{abd}f^{d}f^{c}+f^{a}\epsilon^{bcd}f^{d}],\\
T^{abc}&=&\frac{1}{(1+\vec{f}\cdot\vec{f})^{2}}[\epsilon^{abc}+(1+\vec{f}\cdot\vec{f})f^{b}\delta^{ac}-(1+\vec{f}\cdot\vec{f})f^{c}\delta^{ab}\nonumber\\
&+&\epsilon^{abd}f^{d}f^{c}+f^{a}\epsilon^{bcd}f^{d}+\epsilon^{adc}f^{d}f^{b}].
\end{eqnarray}
The derivation of equations (11) to (14) is straightforward but rather tedious.\\

Note that in the Coulomb gauge limit $(f^{a}=0)$, $t^{a}_{\mu}=A^{a}_{\mu}$ and equations (2) to (14) consistently reduce to the Coulomb gauge results.\\

The pure f dynamics suggests non-perturbative physics because the kinetic term is $(\partial f)^{4}$ and its action is $\sim\frac{1}{g^{2}}$ and infinitely non-linear.  This intuition was verified in reference (4) where it was shown that a stochastic treatment of a class of classical configurations, i.e., those that are spherically symmetric in 4D Euclidean space-time leads to the linear potential and the area law behaviour of the Wilson loop.  Full quantum treatment of the scalar fields, on the other hand, leads to dimensional reduction as the scalar dynamics was shown to be equivalent to a 2D O(1,3) non-linear sigma model via the Parisi-Sourlas mechanism.\\

Before we proceed in the next section with the discussion on the effective dynamics of the ``gluon'' field (the quotes allude to the fact that we deal with SU(2) and refer to $t^{a}_{\mu}$), let us point out features of the pure f dynamics.\\

The class of classical configurations, the spherically symmetric configurations $\tilde{f}^{a}(x)$, not only satisfies
\begin{equation}\label{9}
\frac{\delta S_{f}}{\delta f^{a}(x)}\vert_{\tilde{f}}=0,
\end{equation}
but also
\begin{equation}\label{10}
Z^{a}_{\mu\nu}(\tilde{f}) = 0,
\end{equation}
where
\begin{equation}\label{11}
S_{f} = \int d^{4}x\frac{1}{4}Z^{a}_{\mu\nu} Z^{a}_{\mu\nu}.
\end{equation}
Equations (l5,16) follow from the fact that for spherically symmetric fields $\partial_{\mu}\tilde{f}^{a} = \frac{x_{\mu}}{x}\frac{d\tilde{f}^{a}}{dx}$ and $X^{abc}$ is anti-symmetric with respect to b and c.  This means that $S_{f}$ which is positive definite, has a broad minimum in the function space of the scalars.  It also means that all the spherically symmetric scalars are elements of the classical vacuum.  It is not apparent if the spherically symmetric functions exhaust all classical configurations or the classical vacuum.\\

What is intriguing about the result of reference (4) is that the linear potential follows from a stochastic treatment of vacuum configurations with zero field strength $Z^{a}_{\mu\nu}$.  Is there a simple way to understand this result?  One argument is that the stochastic vacuum configurations  reduce the dimension of space-time from 4 to 2 resulting in a linear potential.  This was confirmed in reference (6) where the full quantum dynamics of $f^{a}$ was shown to be equivalent to a 2D O(1,3) non-linear sigma model.\\

\section{The ``gluon'' in the vacuum defined by scalars}

In equation (2), we will consider the ``gluon'' $t^{a}_{\mu}$ in the vacuum defined by the spherically symmetric scalars $\tilde{f}^{a}(x)$, where $x=(x\cdot x)^{\frac{1}{2}}$.  The resulting expansion is not the usual background decomposition given by
\begin{equation}\label{12}
A^{a}_{\mu} = \tilde{A}^{a}_{\mu}[\tilde{f}]+t^{a}_{\mu},
\end{equation}
as can be seen from equation (2) because $t^{a}_{\mu}$ is coupled in a non-trivial way with $\tilde{f}^{a}(x)$.  Equation (3) was used in reference (5) to derive the linear potential by assuming a white-noise distribution for the spherically symmetric $\tilde{f}^{a}(x)$.  Equation (4) on the other hand, will lead to the gauge condition
\begin{equation}\label{13}
x_{\mu}t^{a}_{\mu} = 0,
\end{equation}
which is known in the literature as the radial gauge\cite{linear},\cite{string}.

Using equations (7) to (14), we find that the effective dynamics
of $t^{a}_{\mu}$ in the vacuum defined by $\tilde{f}^{a}(x)$ is
given by the action
\begin{eqnarray}\label{14}
S_{eff}(t^{a}_{\mu}, \tilde{f}^{a}) &=&\frac{1}{4}\int d^{4}x\{\mathbb{R}^{ab}(\tilde{f})(\partial_{\mu}t^{a}_{\nu}-\partial_{\nu}t^{a}_{\mu})(\partial_{\mu}t^{b}_{\nu}-\partial_{\nu}t^{a}_{\mu})\nonumber\\
&+& 2\mathbb{S}^{ab}(\tilde{f})(\partial_{\mu}t^{a}_{\nu}-\partial_{\nu}t^{a}_{\mu})(\frac{x_{\mu}}{x}t^{b}_{\nu}-\frac{x_{\nu}}{x}t^{b}_{\mu})\nonumber\\
&+&2\mathbb{Y}^{ab}(\tilde{f})t^{a}_{\mu}t^{b}_{\mu}+2g\mathcal{\mathbb{U}}^{abc}(\partial_{\mu}t^{a}_{\nu}-\partial_{\nu}t^{a}_{\mu})t^{b}_{\mu}t^{c}_{\nu}\nonumber\\
&+&g^{2}\mathbb{T}^{abcd}(\tilde{f})t^{a}_{\mu}t^{b}_{\nu}t^{c}_{\mu}t^{d}_{\nu}\},
\end{eqnarray}
where we used equation (19). The coefficients are given by
\begin{eqnarray}\label{14.1}
\mathbb{R}^{ab}(\tilde{f}) &=& R^{ca}(\tilde{f})R^{cb}(\tilde{f}),\\
\mathbb{S}^{ab}(\tilde{f}) &=& R^{ca}Y^{cdb}\frac{d\tilde{f}^{d}}{dx},\\
\mathbb{Y}^{ab}(\tilde{f}) &=& Y^{cda}(\tilde{f})Y^{ceb}(\tilde{f})\frac{df^{d}}{dx}\frac{df^{e}}{dx},\\
\mathbb{U}^{abc}(\tilde{f}) &=& R^{da}(\tilde{f})T^{dbc}(\tilde{f}),\\
\mathbb{T}^{abcd} &=& T^{eab}(\tilde{f})T^{ecd}(\tilde{f}).
\end{eqnarray}

From equation (20), the equation of motion for $t^{a}_{\mu}$ is:
\begin{eqnarray}\label{15}
\frac{\delta S_{eff}}{\delta t^{a}_{\mu}(x)}&=& \mathbb{R}^{ab}(\tilde{f})\Box^{2}t^{b}_{\mu}+\mathbb{M}^{ab}(\tilde{f})t^{b}_{\mu}+\mathbb{N}^{ab}_{\mu\nu}(\tilde{f},x,\partial)t^{b}_{\nu}+\mathbb{P}^{a}_{\mu}(\tilde{f},x)\nonumber\\
&+& g\mathbb{U}^{abc}(\tilde{f})\partial_{\nu}t^{b}_{\mu}t^{c}_{\nu}-\frac{g}{2}\mathbb{U}^{abc}(\tilde{f})(\partial_{\mu}t^{b}_{\nu})t^{c}_{\nu}\nonumber\\
&+&\frac{g^{2}}{2}\mathbb{T}^{abcd}t^{b}_{\nu}(t^{c}_{\mu}t^{d}_{\nu}-t^{c}_{\nu}t^{d}_{\mu}) = 0,
\end{eqnarray}
where
\begin{eqnarray}\label{16}
\mathbb{M}^{ab}(\tilde{f}) &=& \mathbb{Y}^{ab}+\frac{1}{\ell^{2}}U^{abc}\tilde{f}^{c}-\frac{1}{2}\frac{\delta \mathbb{S}^{ab}}{\delta\tilde{f}^{d}}\frac{d\tilde{f}^{d}}{dx},\\
\mathbb{N}^{ab}_{\mu\nu}(\tilde{f},x,\partial) &=& \frac{\delta\mathbb{R}^{ab}}{\delta\tilde{f}^{c}}\frac{d\tilde{f}^{c}}{dx}(\frac{1}{x})[x\cdot\partial+1]\delta_{\mu\nu}+\frac{1}{2}(\mathbb{S}^{ba}-\mathbb{S}^{ab})(\frac{1}{x})\delta_{\mu\nu}(x\cdot\partial)\nonumber\\
&+&(\frac{1}{2}\mathbb{S}^{ba}-\mathbb{S}^{ab})(\frac{1}{x})\delta_{\mu\nu},\\
\mathbb{P}^{a}_{\mu}(\tilde{f}) &=& -\frac{1}{g\ell^{2}}(\frac{x_{\mu}}{x})\mathbb{R}^{ab}\frac{d\tilde{f}^{ab}}{dx}+\frac{1}{2}(\frac{1}{g\ell^{2}})(\frac{x_{\mu}}{x})\mathbb{S}^{ab}\tilde{f}^{b}.
\end{eqnarray}
To derive equations (26) to (29), integration by parts and equation (3) were used.  The last three terms are interaction terms of the vector fields in the presence of the $\tilde{f}$.  The first three terms give the propagation of the ``gluon'' in the vacuum defined by $\tilde{f}$.  The fourth is $t^{a}_{\mu}$ independent and acts like a source term.  Appendix A provides the missing steps in deriving equations (20) to (29).\\

Note the dynamics break translation invariance because of the explicit appearance of $x_{\mu}$.  This is due to the fact that the classical configurations $\tilde{f}^{a}(x)$ implicitly assumes a center, the origin.  However, this is not a concern because when the averaging over $\tilde{f}^{a}(x)$ is done, translation invariance will be recovered.

At this point, the dynamics seem unwieldy.  However, since all spherically symmetric $\tilde{f}^{a}(x)$ are vacuum solutions, we should average over all these configurations.  As in reference (4), we use the white-noise distribution and compute
\begin{equation}\label{17}
\mathcal{N}^{-1}\int(d\tilde{f}^{a}(x))\frac{\delta S_{eff}}{\delta t^{a}_{\mu}(x)}e^{-\frac{1}{\ell}\int^{\infty}_{0}ds\tilde{f}^{a}(s)\tilde{f}^{a}(s)} = 0.
\end{equation}
where $\ell$ was introduced for dimensional reasons.\\

The question now is why assume a white-noise distribution for the spherically symmetric $\tilde{f}^{a}(x)$?  One reason is that it works, i.e., it yielded the area law behaviour for the Wilson loop (see ref. 4).  Aside from this, it also means that the ''gluon" $t^{a}_{\mu}$ is subjected to a random force, albeit, a rather complex combination of white noises.  What we have to do is to find out if this assumption leads to a testable prediction and the answer is yes and this is the subject of the succeeding computations.
We will discretize the integral in the exponent by
\begin{equation}\label{18}
\int^{\infty}_{0}ds\tilde{f}^{a}(s)\tilde{f}^{a}(s) = \sum_{s}\tilde{f}^{a}(s)\tilde{f}^{a}(s)\Delta s.
\end{equation}
For each s, the integration in $\tilde{f}(s)$ becomes ordinary integrals involving
\begin{eqnarray}\label{19}
K^{(4)} &=& N^{-1}\int^{\infty}_{0}r^{4}e^{-\sigma r^{2}}=\frac{3}{2}(\pi\sigma)^{-1},\\
I^{(n,m)} &=& N^{-1}\int^{\infty}_{0}\frac{(r^{2})^{m}}{(1+r^{2})^{n}}e^{-\sigma r^{2}}dr,
\end{eqnarray}
where m and n are integers, $\sigma = \frac{\Delta s}{\ell}$ and $N = \pi^{\frac{3}{2}}\sigma^{-\frac{3}{2}}$ is the normalization factor, the products of which for all s values give the overall normalization $\mathcal{N}$ found in equation (30).  Equation (33) is evaluated using integrals derived from (by differentiating w.r.t. $\beta$ or $\sigma$)
\begin{equation}\label{20}
\int^{\infty}_{0}\frac{e^{-\sigma r^{2}}}{(r^{2}+\beta^{2})}dr = [1-\Phi(\sigma^{\frac{1}{2}}\beta)\frac{\pi}{2\beta}e^{\beta^{2}\sigma}],
\end{equation}
where $Re\beta > 0$ and arg $\sigma<\frac{\pi}{4}$ (see reference\cite{integral}). The function $\Phi$ is the error function given by
\begin{eqnarray}\label{21}
\Phi(\sigma^{\frac{1}{2}}\beta) &=& \frac{2}{\sqrt{\pi}}\int^{\beta\sigma\frac{1}{2}}_{0}e^{-t^{2}}dt,\nonumber\\
&=&\frac{2}{\sqrt{\pi}}[\beta\sigma^{\frac{1}{2}}-\frac{1}{3}(\beta\sigma^{\frac{1}{2}})^{3}+\frac{1}{10}(\beta\sigma^{\frac{1}{2}})^{5}-\frac{1}{42}(\beta\sigma^{\frac{1}{2}})^{7}+...].
\end{eqnarray}

The terms to average stochastically are given by equations (24) to (29).  There are three types depending on the number of derivatives of $\tilde{f}^{a}$.  The terms without derivatives are $\mathbb{R}, \mathbb{U}$ and $\mathbb{T}$ given by equations (21), (24), (25), (12) and (14).  The results are
\begin{eqnarray}\label{22}
\langle\mathbb{R}^{ab}\rangle_{\tilde{f}}&=&\frac{1}{3}\delta^{ab},\\
\langle\mathbb{U}^{abc}\rangle_{\tilde{f}}&=&\langle\mathbb{U}^{abc}\tilde{f}^{c}\rangle_{\tilde{f}}=\langle\mathbb{T}^{abcd}\rangle_{\tilde{f}}=0.
\end{eqnarray}

The second and third types of terms contain $\frac{d\tilde{f}^{a}}{dx}$ and $\frac{d\tilde{f}^{a}}{dx}\frac{d\tilde{f}^{b}}{dx}$ respectively.  Thus, there is a need to define the derivative of a white noise.  Since a white noise is continuous but not differentiable, i.e., the limit of $\frac{\tilde{f}^{a}(x+\Delta x)-\tilde{f}^{a}(x)}{\Delta x}$ as $\Delta x \to 0$ does not exist, we need to define $\frac{d\tilde{f}^{a}}{dx}$.  This will be done by a suitable smoothening process.  Since there is already a natural length scale $\ell$ in this regime, we will define
\begin{equation}\label{22.1}
\frac{d\tilde{f}^{a}}{dx}=\frac{\tilde{f}^{a}(x+\frac{\ell}{n})-\tilde{f}^{a}(x)}{\ell/n},
\end{equation}
where $n$ is a number $\geq 1$.  Obviously, we cannot take $n\to\infty$, because of the white-noise behaviour.\\

Using equation (38), all the terms with one derivative, vanish.
\begin{equation}\label{23}
\langle\frac{\delta\mathbb{R}^{ab}}{\delta\tilde{f}^{c}}\frac{d\tilde{f}^{c}}{dx}\rangle_{\tilde{f}}=\langle\mathbb{S}^{ab}\rangle_{\tilde{f}}=0.
\end{equation}
This in turn yields
\begin{equation}\label{24}
\langle\mathbb{N}^{ab}_{\mu\nu}\rangle_{\tilde{f}}=\langle\mathbb{P}^{a}_{\mu}\rangle_{\tilde{f}}=0.\\
\end{equation}

There are two terms with two derivatives and these are
\begin{eqnarray}\label{25}
\langle\frac{\delta\mathbb{S}^{ab}}{\delta\tilde{f}^{c}}\frac{d\tilde{f}^{d}}{dx}\rangle_{\tilde{f}}=0\\
\langle\mathbb{Y}^{ab}\rangle_{\tilde{f}}=\delta^{ab}(\frac{3}{4})\frac{n^{2}}{\ell^{2}}.
\end{eqnarray}
The details of all these computation are shown in Appendix B.\\

All these simplify equations (24) and (30) to yield
\begin{equation}\label{26}
(\Box^{2} + \frac{9}{4} \frac{n^{2}}{\ell^{2}})t^{a}_{\mu}=0.
\end{equation}
This means the effective classical dynamics of the gluons show that they are non-interacting and have a range of possible masses given by
\begin{equation}\label{27}
m_{g}=\frac{3}{2}\cdot\frac{n}{\ell},\quad   n\geq 1.
\end{equation}
At the very least this result verifies the existence of the mass gap for non-Abelian gauge theories because the smallest value of the ''gluon's" $(t^{a}_{\mu})$ mass is $\frac{3}{2}(\frac{1}{\ell})=\frac{3}{2}\Lambda_{QCD}$.  In the perturbative regime, the gluon $A^{a}_{\mu}$ is massless.\\

Is n an integer?  Typically, we subdivide a length (in this case the confinement length scale $\ell=\frac{1}{\Lambda_{QCD}}$ into a finite number of equally spaced segments.  Experimentally, the way to determine $n$ is to compare the predicted mass with the mass spectrum of massive spin one ''pure glue states".  The lowest mass, which gives the mass gap, will determine $n_{min}$ while the other states will determine the other values of $n$. Unfortunately, we cannot do this comparison because the computation was done for $SU(2)$ and not $SU(3)$ (see Appendix 2).  But the result will essentially be the same, i.e., $m\sim n\Lambda_{QCD}$, only the proportionality factor $\frac{3}{2}$ will change.\\

To conclude this section, we comment on the seemingly paradoxical result of the gluon $t^{a}_{\mu}$ acquiring mass and classically losing self-interaction.  What was shown is that the effective classical dynamics of $t^{a}_{\mu}$ in the stochastic vacuum defined by the spherically symmetric $\tilde{f}^{a}(x)$ gives a non-interacting and massive $t^{a}_{\mu}$.  This is not inconsistent with confinement because the quantum dynamics of $t^{a}_{\mu}$ in a particular classical background $\tilde{f}^{a}(x)$ necessitates the evaluation of the $t^{a}_{\mu}$ propagator given by

\begin{eqnarray}\label{25.1}
\left[\mathbb{R}^{ab}(\tilde{f})\Box^{2}+\mathbb{M}^{ab}(\tilde{f})\right]G^{bc}(x,y;\tilde{f})=\delta^{ac}\delta^{4}(x-y).
\end{eqnarray}
After computing $\tilde{G}(x,y;\tilde{f})$, we average over $\tilde{f}^{a}(x)$ to get the stochastic average of the propagator.  This is certainly not equivalent to the propagator of the stochastically averaged $\langle \mathbb{R}^{ab}\Box^{2} +\mathbb{M}^{ab}\rangle_{\tilde{f}}$ as can be seen from equation (45).  In other words, stochastic averaging does not factorize.

\section{Limited Gauge-Invariance}

At this point, we speculate on the gauge-invariance of the result.  The claim is that the linear and non-linear regimes of the non-linear gauge satisfy \underline{limited} gauge invariance.  To explain what is meant by this, consider first an Abelian theory.  Starting from a field $A_{\mu}$ that does not satisfy the Coulomb gauge, one can project to the Coulomb surface to get $\tilde{A}_{\mu}$ or do a gauge transformation to get $A'_{\mu}$, which satisfies the Coulomb gauge.  The results in both cases are the same, i.e.,
\begin{equation}\label{26.1}
A'_{\mu}=\tilde{A}_{\mu}=(\delta_{\mu\nu}-\partial_{\mu}\frac{1}{\Box^{2}}\partial_{\nu})A_{\nu}.
\end{equation}
Furthermore, the transverse field configuration $A'_{\mu}$(and $\tilde{A}_{\mu})$ is also gauge-invariant.\\

In the non-Abelian case, the situation is more tricky.  First, the results of the projection (similar to equation (46) only with an additional SU(2) index $a$) and the gauge transformation to the Coulomb surface, (if it exists) are not the same.  Second, the projection to the Coulomb surface is not gauge-invariant.\\

To illustrate limited gauge-invariance, consider a field configuration $A^{a}_{\mu}$ such that $\partial\cdot A^{a}=\epsilon^{a}(x)$ where $\epsilon^{a}(x)$ is infinitesimal and the operator $(\partial\cdot D)$ positive definite.  An infinitesimal gauge transformation that will take this field configuration to the Coulomb gauge exists and it is
\begin{equation}\label{27.1}
A'^{a}_{\mu}=A^{a}_{\mu}-D^{ab}_{\mu}(x)\int d^{4}y G^{bc}(x;y;A)(\partial\cdot A^{c})_{y},
\end{equation}
where $G^{ab}$ is the Green function of $(\partial\cdot D)$.  Note that if $(\partial\cdot D)$ has $L^{2}$ zero mode $z^{a}(x;A)$, then the solution given by equation (47) is not valid.  Strictly speaking, we need to make use of the Green function
\begin{equation}\label{28}
G'^{ab}=G^{ab}-z^{a}(x)z^{b}(y),
\end{equation}
in equation (47) and also impose that $\int d^{4}x z^{a}(\partial\cdot A^{a}(x)) = 0$.  This is the reason why we require the positive-definiteness of $(\partial\cdot D)$.\\

The field configuration given by equation (47) is transverse.  It is also gauge-invariant as we will show.  First $\delta A'^{a}_{\mu} = D^{ab}_{\mu}(A')\sigma^{b}$.  The right hand side of equation (47) on the other hand can be evaluated by using $\delta A^{a}_{\mu}=D^{ab}_{\mu}\Lambda^{b}$ and
\begin{equation}\label{29}
\delta G^{ab}(x,y;A)= -\int d^{4}z G^{ac}(x,z;A)\delta(\partial\cdot D)^{cd}_{z}G^{db}(z;y;A).
\end{equation}
However, this is too complicated to evaluate.  Instead acting by $\partial_{\mu}$ on both sides of the variation of equation (47), it is trivial to show that the RHS yield exactly zero\cite{magpantay}.  Thus, we have
\begin{equation}\label{30}
(\partial\cdot D^{ab}(A'))\sigma^{b}=0.
\end{equation}
Second, since $det(\partial\cdot D)$ is invariant under infinitesimal transformation and since $\partial\cdot D(A)$ is positive-definite, then $det(\partial\cdot D(A'))$ is also non-zero and $\partial\cdot D(A')$ is also non-singular.  Thus, $\sigma^{a}=0$, proving that $A'^{a}_{\mu}$ given by equation (47) is gauge-invariant.  Since equation (47) is only defined for $(\partial\cdot D)$ positive-definite, then we were only able to derive a transverse, gauge-invariant potential subject to this condition.\\

Now consider field configurations $A^{a}_{\mu}(x)$ such that $\partial\cdot A^{a}=f^{a}(x)$ and $(\partial\cdot D)^{ab}f^{b}=\epsilon^{a}(x)\neq 0$, where
$\epsilon^{a}(x)$ is an infinitesimal function.  Assuming that this field configuration can be gauge transformed to the non-linear regime of the non-linear gauge (and as reference (2) showed it cannot be gauge transformed to the Coulomb gauge), i.e., $A'^{a}_{\mu}=A^{a}_{\mu}+D^{ab}_{\mu}(A)\rho^{b}$, such that $\partial\cdot A'^{a}\neq 0$ and $(\partial\cdot D^{ab}(A'))(\partial\cdot A'^{b})=0$.  In this case,
\begin{equation}\label{31}
\rho^{a} = -\int d^{4}y H^{ab}(x,y;A)[(\partial\cdot D^{bc}(A))(\partial\cdot A^{c})]_{y},
\end{equation}
where $H^{ab}$ is the Green function of the fourth-order operator
\begin{equation}\label{32}
\theta^{ab}=(D\cdot\partial)^{ac}(\partial\cdot D)^{cb}-\epsilon^{acd}[\partial(\partial\cdot A^{d})]\cdot D^{cb}.
\end{equation}
The gauge-potential $A'^{a}_{\mu}$ given by
\begin{equation}\label{33}
A'^{a}_{\mu}=A^{a}_{\mu}-D^{ab}_{\mu}\int d^{4}y H^{bc}(x,y;A)[(\partial\cdot D)^{cd}(\partial\cdot A^{d})]_{y}
\end{equation}
is gauge-invariant.  The proof is essentially the same as in the Coulomb gauge only the steps are more involved.\\

Under gauge transformation, $\delta A'^{b}=D^{bc}_{\mu}(A')\sigma^{c}$, while $\delta A^{b}_{\mu}=D^{bc}_{\mu}(A)\Lambda^{c}$.  Then we act $\partial_{\mu}$ on both sides of the gauge variations of equation (53) and followed by $(D^{ab}_{\mu}(A')\cdot\partial)$.  From this result, we subtract $\epsilon^{abc}[\partial_{\mu}(\partial\cdot A^{\prime})]\delta A'^{c}_{\mu}$.  The left-hand side is simply $\theta^{ab}(A^{\prime})\sigma^{b}$.\\

The right-hand side is rather involved because $\delta A^{a}_{\mu}$ involves $\Lambda^{b}$ while the operator that we used involved $A'^{a}_{\mu}$.  Careful evaluation of these terms yield zero\cite{magpantay} and $\sigma^{a}$ is then determined from
\begin{equation}\label{34}
\theta^{ab}(A')\sigma^{b}=0.
\end{equation}
In reference (12), it was shown in detail that even though the operator $\partial\cdot D(A')$ is singular with $\partial\cdot A'$ as its zero mode, the operator $\theta^{ab}(A')$ is non-singular.  Therefore, the only $L^{2}$ solution of equation (54) is the trivial solution and thus proving the gauge-invariance of the potential $A'^{a}_{\mu}$ given by equation (53).\\

Some details of the above computations are given in Appendix C.\\

We have shown then that the potentials given by equations (47) and (53) are gauge-invariant subject to their respective defining conditions.  Outside these conditions, these potentials are not gauge-invariant in general.  Thus, we say that these potentials respect limited gauge-invariance.\\

The unrestricted gauge-invariance of the Abelian potential defined by equation (46) and the restricted gauge-invariance of the non-Abelian potentials defined by equations (47) and (53) are reflections of the fact that the gauge conditions used in deriving these potentials remove the excess degrees of freedom.  These follow from the fact the relevant operators - $\Box^{2}$ in the Coulomb gauge of the Abelian theory, $(\partial\cdot D)^{ab}$ in the Coulomb gauge of the non-Abelian theory and $\theta^{ab}$ in the non-linear gauge of the non-Abelian theory-are non-singular operators.  In the case of $\Box^{2}$ it is non-singular without restrictions while $(\partial\cdot D)^{ab}$ and $\theta^{ab}$ are only non-singular within limited regimes.  In the Abelian theory, if $A_{\mu}$ is on the Coulomb surface, then $A_{\mu}+\partial_{\mu}\Lambda$ will always be off the Coulomb surface because $\Box^{2}\Lambda\neq 0$ (unless $\Lambda = 0$).  In the non-Abelian case, the non-singularness of $(\partial\cdot D)^{ab}$ and $\theta^{ab}$ in their respective regimes guarantee that the gauge transformed fields in the Coulomb gauge and non-linear gauge respectively would no longer satisfy the gauge condition.  As a consequence, the decomposition given by equation (2), which is only valid if equation (1) is satisfied, is not valid for the gauge transformed field.  Thus, it does not make sense to ask how $f^{a}$ and $t^{a}_{\mu}$ transform under gauge transformation.\\

Equation (47) defines transverse vector fields, which can be viewed as the linear regime of the non-linear gauge.  This is valid when the operator $\partial\cdot D^{ab}=\Box^{2}\delta^{ab}-g\epsilon^{abc}\partial\cdot A^{c}$ is positive definite.  This happens when the second term is not strong enough to lower the lowest eigenstate of $\Box^{2}$ to reach the zero eigenvalue level.  And this is true in the perturbative regime where $g$ is rather weak and $A^{a}_{\mu}$ is a fluctuation off the trivial vacuum $A^{a}_{\mu}=0$.\\

As the coupling increases in strength (increasing distance scale), then the lowest mode of $\Box^{2}$ can be lowered down to the zero eigenvalue level.  We now hit the Gribov horizon where the zero mode of $(\partial\cdot D)$ is $z^{a}(x)$.  $A^{a}_{\mu}$ still represents fluctuations off the trivial vacuum.  Here, we cannot define a gauge-invariant potential.  There are some claims that non perturbative physics is accounted for by this regime.  We take a differing opinion because as argued in reference (1), these field configurations do not contribute to the path-integral.\\

As we increase further the distance scale, resulting in increasing $g$, there comes a point when $(\partial\cdot D)$ is singular and $\partial\cdot A^{a}$ itself is its zero mode.  We now hit the Gribov horizon off the Coulomb surface.  This is the non-linear regime of the non-linear gauge.  In this regime, we can define the gauge-invariant potential given by equation (53).  Also in this regime, the decomposition of the potential in terms of $f^{a}$ and $t^{a}_{\mu}$ given by equation (2) is valid.  The set of vacuum configuration is very degenerate, it includes all spherically symmetric $\tilde{f}^{a}(x)$.  The stochastic dynamics of $\tilde{f}^{a}(x)$ yields the linear potential and generates mass for the gluon $t^{a}_{\mu}$.  The full dynamics of $f^{a}(x)$ on the other hand, give the Parisi-Sourlas mechanism.  Although these are derived from the non-linear regime of the non-linear gauge, we argue that these are gauge-invariant results because a gauge-invariant potential (given by equation (53)) can be defined in that regime.

\section{Conclusion}

In this paper, we argued that the class of spherically symmetric $\tilde{f}^{a}(x)$ (in $R^{4}$) form vacuum configurations with zero action.  Then it was shown that by treating $\tilde {f}^{a}(x)$ as stochastic, the effective dynamics of the ``gluons'' in this background result in the vanishing of the ``gluon'' self-interaction and generation of its mass.\\

It is apparent that the class of spherically-symmetric zero field strength $(Z^{a}_{\mu\nu}=0)$ vacuum configurations that produced the linear potential and ``gluon'' mass is in disagreement with current ideas on the nature of the QCD vacuum.  In the 1970's and early 80's, Pagels and Tomboulis\cite{Pagels}, and Savvidy\cite{Sav} showed that a non-zero field strength configuration has lower energy than the zero field strength configuration.  Unfortunately, the effective Hamiltonian has an imaginary part which signals instability and eventual decay to the trivial vacuum.\\

About the same time the concept of condensates\cite{Shifman}, (in particular, the $\langle 0\mid F^{2}\mid 0\rangle\neq 0)$, supports the idea of a vacuum with non-zero field strength.  Obviously, the ideas presented here and the condensate result are contradictory.  Or are they?\\

Note that the decomposition of the original $A^{a}_{\mu}$ given in equation (2) shows the scalar field $f^{a}$ inextricably linked to the $t^{a}_{\mu}$ and not in the usual background decomposition $A^{a}_{\mu} = \tilde {A}^{a}_{\mu} + t^{a}_{\mu}$.  When we expand the field strength, we get equation (7), which shows $t^{a}_{\mu}$ interacting with $f$ in a non-linear manner.  The vacuum configurations defined by $f^{a}$, which are the class of spherically symmetric configurations $\tilde{f}^{a}(x)$, have vanishing $Z^{a}_{\mu\nu}$.  The non-zero condensate $\langle 0\vert F^{2}\vert 0\rangle$ involves $L^{a}_{\mu\nu}$ and $Q^{a}_{\mu\nu}$, i.e., it also reflects the $t^{a}_{\mu}$ dynamics.  Thus, the ideas presented here and that of the non-vanishing condensate are not necessarily contradictory.\\

We also argued the gauge-invariance of the results derived.  This is done by showing that a gauge-invariant potential, which satisfies the non-linear gauge can be defined within a limited regime of potentials.\\

Finally, we make the following observation about the non-linear gauge.  The non-linear gauge condition, as already stated, interpolates between the Coulomb gauge (with weakly interacting transverse gluons at the short-distance regime) and the quadratic regime (with $f^{a}$ and $t^{a}_{\mu}$ as degrees of freedom, which accounts for confinement and the mass for the vector field in the long-distance regime).  This assumes that the coupling runs, implying that the non-linear gauge is essentially a quantum gauge condition.  This is an unusual way to fix the gauge.  Normally, the gauge condition represents a fixed submanifold in configuration space regardless of the distance scale of the physics.  The non-linear gauge, on the other hand, defines a quadratic submanifold, which, depending on the coupling, will highlight either the transverse degrees of freedom of the massless vector field (in the linear or Coulomb regime) or the scalar $f^{a}$ and the massive vector field $t^{a}_{\mu}$ (in the non-linear regime).\\

\section{Acknowledgement}
This research was supported in part by the Natural Sciences Research Institute and by the University of the Philippines System.  I would like to thank Pol Nazarea for enlightening me on the nature of white-noise.\\


\newpage
\appendix

\section{Derivation of the action and field equations}

From equations (7) to (10), we find
\newcommand{\onefourth}{\frac{1}{4}}
\begin{eqnarray}\label{A.1}
\mathcal{L}_{YM}
    & = & \onefourth F^{a}_{\mu\nu}F^{a}_{\mu\nu}\nonumber\\
    & = & \onefourth \left\{ R^{ab}
            ( \partial_{\mu}t^{b}_\nu - \partial_\nu t^b_\mu )
            R^{ac} ( \partial_\mu t^c_\nu - \partial_\nu t^c_\mu ) \right.\nonumber\\
   &  & +  Y^{abc} ( \partial_\mu \tilde{f}^b t^c_\nu - \partial_\nu \tilde{f}^b t^c_\mu )
                Y^{ade} ( \partial_\mu \tilde{f}^d t^e_\nu - \partial_\nu \tilde{f}^d t^e_\mu )\nonumber \\
   &  & + 2 R^{ab} ( \partial_{\mu}t^{b}_\nu - \partial_\nu t^b_\mu )
               Y^{acd} ( \partial_\mu \tilde{f}^c t^d_\nu - \partial_\nu \tilde{f}^c t^d_\mu )\nonumber\\
   &  & + 2 gR^{ab}(\partial_{\mu} t^{b}_{\nu} - \partial_\nu t^{b}_{\mu})T^{acd}t^{c}_{\mu}\nonumber\\
   &  & + 2 g Y^{abc}(\partial_{\mu}\tilde{f}^{b}t^{c}_{\nu}-\partial_{\nu}\tilde{f}^{b}t^{c}_{\mu})T^{ade}t^{d}_{\mu}t^{e}_{\nu}\nonumber\\
   &  & + g^{2}T^{abc}t^{b}_{\mu}t^{c}_{\nu}T^{ade}t^{d}_{\mu}t^{e}_{\nu}\left. \right\}
\end{eqnarray}
Using $\partial_{\mu}\tilde{f}^{b} = \frac{x_{\mu}}{x}\frac{d\tilde{f}^{b}}{dx}$, and equation (19), we find that the $Y\cdot T$ term vanishes.  Also, the $Y\cdot Y$ term becomes
\begin{eqnarray}\label{A.2}
(Y\cdot Y)_{term}=2
Y^{abc}Y^{ade}\frac{d\tilde{f}^{b}}{dx}\frac{d\tilde{f}^{d}}{dx}t^{c}_{\mu}t^{e}_{\mu}
\end{eqnarray}
Substituting these results give the action shown in equations (20) to (25).\\

Equation (20) gives the action of the ``gluon" $t^{a}_{\mu}$ in a classical background $\tilde{f}^{a}(x)$.  In deriving the field equation equation given by equations (26) to (29); we made use of integration by parts and equations (3) and (19).  For example, the variation of the first and second terms of equation (20) gives
\begin{eqnarray}
first\ term &=& \mathbb{R}^{ab}\Box^{2}t^{b}_{\mu}-\frac{1}{g\ell^{2}}\mathbb{R}^{ab}\frac{x_{\mu}}{x}\frac{d\tilde{f}^{b}}{dx}\nonumber\\
&+& \frac{\delta\mathbb{R}^{ab}}{\delta\tilde{f}^{c}}\frac{d\tilde{f}^{c}}{dx}(\frac{1}{x})(x\cdot \partial)t^{b}_{\mu}\nonumber\\
&+& \frac{\delta\mathbb{R}^{ab}}{\delta f^{c}}\frac{d\tilde{f}^{c}}{dx}(\frac{1}{x})t^{b}_{\mu}\label{A.3}\\
second\ term&=&\frac{1}{2}(\mathbb{S}^{ba}-\mathbb{S}^{ab})(\frac{1}{x})(x\cdot\partial)t^{b}_{\mu}+(\frac{1}{2}\mathbb{S}^{ba}-\mathbb{S}^{ab})(\frac{1}{x})t^{b}_{\mu}\nonumber\\
&+&\frac{1}{2}\mathbb{S}^{ab}(\frac{x_{\mu}}{x})\frac{1}{g\ell^{2}}\tilde{f}^{b}-\frac{1}{2}\frac{\delta\mathbb{S}^{ab}}{\delta\tilde{f}^{d}}\frac{d\tilde{f}^{d}}{dx}t^{b}_{\mu}\label{A.4}
\end{eqnarray}

\section{Stochastic Averages}

We need to evaluate the stochastic averages of the following terms:
\begin{eqnarray}
\mathbb{R}^{ab}&=&\frac{1}{(1+\vec{f}\cdot\vec{f})}(\delta^{ab}+f^{a}f^{b})\label{B.1}\\
\mathbb{Y}^{ab}&=&\frac{1}{(1+\vec{f}\cdot\vec{f})}+\left\{(\vec{f}\cdot\vec{f})^{2}\frac{d\vec{f}^{c}}{dx}\frac{df^{c}}{dx}\delta^{ab}\right.\nonumber\\
&+&[1+\vec{f}\cdot\vec{f}+(\vec{f}\cdot\vec{f})^{2}]f^{c}\frac{df^{c}}{dx}f^{d}\frac{df^{d}}{dx}\delta^{ab}+(\vec{f}\cdot\vec{f})[1+\vec{f}\cdot\vec{f}+(\vec{f}\cdot\vec{f})^{2}]\frac{df^{a}}{dx}\frac{df^{b}}{dx}\nonumber\\
&-& [1+\vec{f}\cdot\vec{f}+(\vec{f}\cdot\vec{f})^{2}](f^{a}\frac{df^{b}}{dx}+\frac{df^{a}}{dx}f^{b})f^{c}\frac{df^{c}}{dx}\nonumber\\
&-& (1-\vec{f}\cdot\vec{f})\epsilon^{acd}f^{c}\frac{df^{d}}{dx}f^{b}f^{e}\frac{df^{e}}{dx}-(\vec{f}\cdot\vec{f})f^{a}f^{b}\frac{df^{c}}{dx}\frac{df^{c}}{dx}\nonumber\\
&-& (\vec{f}\cdot\vec{f})f^{c}\epsilon^{cda}f^{g}\epsilon^{gbe}\frac{df^{d}}{dx}\frac{df^{e}}{dx}-(1-\vec{f}\cdot\vec{f})f^{c}\epsilon^{cae}f^{b}f^{d}\left.\frac{df^{d}}{dx}\frac{df^{e}}{dx}\right\}\label{B.2}
\end{eqnarray}
\begin{eqnarray}
\mathbb{U}^{abc}&=&\frac{1}{(1+\vec{f}\cdot\vec{f})^{3}}\left\{\epsilon^{abc}+(2+\vec{f}\cdot\vec{f})\epsilon^{abd}f^{d}f^{c}\right.\nonumber\\
&+& \left.(2+\vec{f}\cdot\vec{f})f^{a}\epsilon^{bcd}f^{d}+(2+\vec{f}\cdot\vec{f})\epsilon^{adc}f^{d}f^{b}\right\}\label{B.3}
\end{eqnarray}
\begin{eqnarray}
\mathbb{T}^{abcd}&=&\frac{1}{(1+\vec{f}\cdot\vec{f})^{4}}\left\{[\epsilon^{eab}+(1+\vec{f}\cdot\vec{f})f^{a}\delta^{eb}-(1+\vec{f}\cdot\vec{f})f^{b}\delta^{ea}\right.\nonumber\\
&+& \epsilon^{eaf}f^{b}f^{f}+f^{e}\epsilon^{abf}f^{f}+\epsilon^{efb}f^{a}f^{f}][\epsilon^{ecd}+(1+\vec{f}\cdot\vec{f})f^{c}\delta^{ed}\nonumber\\
&-& \left.(1+\vec{f}\cdot\vec{f})f^{d}\delta^{ec}+\epsilon^{ecg}f^{g}f^{d}+f^{e}\epsilon^{cdg}f^{g}+\epsilon^{egd}f^{g}f^{c}]\right\}\label{B.4}
\end{eqnarray}
\begin{eqnarray}
\mathbb{S}^{ab}&=&\frac{1}{(1+\vec{f}\cdot\vec{f})^{3}}\left\{[(\vec{f}\cdot\vec{f})\epsilon^{abc}+(1+\vec{f}\cdot\vec{f})^{2}f^{a}\delta^{bc}-(1+\vec{f}\cdot\vec{f})\delta^{ab}f^{c}\right.\nonumber\\
&-& (2+\vec{f}\cdot\vec{f})\epsilon^{abd}f^{d}f^{c}-(1+\vec{f}\cdot\vec{f})f^{a}f^{b}f^{c}+\epsilon^{acd}f^{b}f^{d}\nonumber\\
&-& \left. f^{a}\epsilon^{bcd}f^{d}]\frac{df^{c}}{dx}\right\}\label{B.5}\
\end{eqnarray}
\begin{eqnarray}
\mathbb{U}^{abc}f^{c}& = &\frac{1}{(1+\vec{f}\cdot\vec{f})}\epsilon^{abc}f^{c}\label{B.6}\
\end{eqnarray}
\begin{eqnarray}
\frac{\delta\mathbb{R}^{ab}}{\delta f^{c}}\frac{df^{c}}{dx}&=&\frac{1}{(1+\vec{f}\cdot\vec{f})^{2}}\left\{[(1+\vec{f}\cdot\vec{f})\delta^{ac}f^{b}+(1+\vec{f}\cdot\vec{f})f^{a}\delta^{bc}\right.\nonumber\\
&-& \left. 2(\delta^{ab}+f^{a}f^{b})f^{c}]\frac{df^{c}}{dx}\right\}\label{B.7}\
\end{eqnarray}
\begin{eqnarray}
\frac{\delta\mathbb{S}^{ab}}{\delta f^{c}}\frac{df^{c}}{dx}&=& \frac{1}{(1+\vec{f}\cdot\vec{f})^{4}}\left\{[-7(\vec{f}\cdot\vec{f})(1+\vec{f}\cdot\vec{f})\epsilon^{abd}f^{c}\right.\nonumber\\
&+&(1+\vec{f}\cdot\vec{f})^{3}\delta^{ac}\delta{bd}-2(1+\vec{f}\cdot\vec{f})^{2}f^{a}f^{c}\delta^{bd}\nonumber\\
&-& (1+\vec{f}\cdot\vec{f})^{2}\delta^{ab}\delta^{cd}+4(1+\vec{f}\cdot\vec{f})\delta^{ab}f^{c}f^{d}\nonumber\\
&+& (2+\vec{f}\cdot\vec{f})(1+\vec{f}\cdot\vec{f})\epsilon^{bae}f^{e}\delta^{cd}-(10+4 \vec{f}\cdot\vec{f})\epsilon^{bae}f^{c}f^{e}f^{d}\nonumber\\
&-& (1+\vec{f}\cdot\vec{f})^{2}\delta^{ac}f^{b}f^{d}-(1+\vec{f}\cdot\vec{f})^{2}\delta^{cd}f^{a}f^{b}\nonumber\\
&+& 4(1+\vec{f}\cdot\vec{f})f^{a}f^{b}f^{c}f^{d}-(1+\vec{f}\cdot\vec{f})^{2}\delta^{bc}f^{a}f^{d}\nonumber\\
&+& (1+\vec{f}\cdot\vec{f})\epsilon^{ade}f^{e}\delta^{bc}+(1+\vec{f}\cdot\vec{f})\delta^{ac}\epsilon^{dbe}f^{e}\nonumber\\
&-& \left. 6\epsilon^{ade}f^{c}f^{e}f^{b}-6\epsilon^{dbe}f^{a}f^{c}f^{e}]\frac{df^{c}}{dx}\frac{df^{d}}{dx}\right\}\label{B.8}
\end{eqnarray}
The average of terms without derivatives (B.1), (B.3), (B.4) and (B.6) are of the form
\begin{eqnarray}
\langle F[f]\rangle_{f}=\mathcal{N}^{-1}\int(df^{a}(x))F[f]e^{-\frac{1}{\ell}\int^{\infty}_{0}f^{a}(x)f^{a}(x)}\label{B.9}\
\end{eqnarray}
Using equation (31) and the fact that $F[f]$ is local in x, the averaging will reduce to ordinary integration
\begin{eqnarray}
\langle F[f]\rangle_{f}=\frac{1}{N}\int r^{2}dr d \Omega F(r)e^{-\sigma r^{2}},\label{B.10}
\end{eqnarray}
where $r^{2}=\vec{f}(x)\cdot\vec{f}(x)$, $\sigma =\frac{\Delta s}{\ell}$, and $N^{-1}=\pi^{-\frac{3}{2}}\sigma^{+\frac{3}{2}}$.  At the end of the computation, $\sigma\rightarrow 0$ must be taken to get the surviving terms.\\
To illustrate the computation, we will show in detail the evaluation of the stochastic average of (B.1).
\begin{eqnarray}
\langle\mathbb{R}^{ab}\rangle_{f}=\pi^{-\frac{3}{2}}\sigma^{+\frac{3}{2}}\int r^{2}dr d\Omega\frac{e^{-\sigma r^{2}}}{(1+r^{2})}(\delta^{ab}+r^{a}r^{b})\label{B.11}
\end{eqnarray}
Using $\int d\Omega r^{a}r^{b} = \frac{4\pi}{3}r^{2}\delta^{ab}$,
\begin{eqnarray}
\langle\mathbb{R}^{ab}\rangle_{f}&=&\pi^{-\frac{3}{2}}\sigma^{+\frac{3}{2}}\left\{4\pi\int^{\infty}_{0}\frac{r^{2}e^{-\sigma r^{2}}}{1+r^{2}}dr\right.\nonumber\\
&+& \left. \frac{4\pi}{3}\int^{\infty}_{0}\frac{r^{4}e^{-\sigma r^{2}}}{(1+r^{2})}dr\right\}\delta^{ab}\nonumber\\
&=& 4\pi^{-\frac{1}{2}}\sigma^{+ \frac{3}{2}}\left\{ \int^{\infty}_{0}e^{-\sigma r^{2}}dr-\int^{\infty}_{0}\frac{e^{-\sigma r^{2}}}{(1+r^{2})}dr\right.\nonumber\\
&+& \left. \frac{1}{3}\int^{\infty}_{0}r^{2}e^{-\sigma r^{2}}dr-\frac{1}{3}\int^{\infty}_{0}e^{-\sigma r^{2}}dr + \frac{1}{3}\int^{\infty}_{0}\frac{e^{-\sigma r^{2}}}{(1+r^{2})}dr\right\}\delta^{ab}\nonumber\\
&=& 4\pi^{-\frac{1}{2}}\sigma^{+\frac{3}{2}}\left\{\frac{2}{3}\int^{\infty}_{0}e^{-\sigma r^{2}}dr + \frac{1}{3}\int^{\infty}_{0}r^{2}e^{-\sigma r^{2}}\right.\nonumber\\
&-& \left. \frac{2}{3}\int^{\infty}_{0}\frac{e^{-\sigma r^{2}}}{(1+r^{2})}dr\right\}\delta^{ab}.\label{B.12}
\end{eqnarray}
The first integral is proportional to $\sigma^{-\frac{1}{2}}$, the third is a series beginning with $\sigma^{+\frac{1}{2}}$ and higher orders (see equations (34) and (35)) while the second is proportional to $\sigma^{-\frac{3}{2}}$.  Thus, in the limit $\sigma\rightarrow 0$, we find
\begin{eqnarray}
\langle\mathbb{R}^{ab}\rangle_{f}=\frac{1}{3}\delta^{ab}\nonumber\
\end{eqnarray}
which is the result given by equation (36).  The averages of (B.3), (B.4) and (B.6), vanish either because (a) they are odd powers in $f^{a}$ or (b) the terms with the highest power of $\sigma$ have powers less than $-\frac{3}{2}$.  For example for (B.4), these are the terms like $(1+\vec{f}\cdot\vec{f})^{2}f^{a}f^{c}\delta^{bd}$, which yield the integral
\begin{eqnarray}
\pi^{-\frac{3}{2}}\sigma^{+\frac{3}{2}}\int^{\infty}_{0}\frac{r^{4}(1+r^{2})^{2}}{(1+r^{2})^{4}}e^{-\sigma r^{2}}.\label{B.13}
\end{eqnarray}
Since both the denominator and numerator are of order $r^{8}$, the most divergent part of the integral is proportional to
\begin{eqnarray}
\int^{\infty}_{0} e^{-\sigma r^{2}}dr \quad  \sim \quad\sigma^{-\frac{1}{2}}.\label{B.14}
\end{eqnarray}
Thus, when we take $\sigma\rightarrow 0$, we get equation (37).\\

For terms with one derivative, which are (B.5) and (B.7), we make use of equation (38).  Now there are two space-time points, at $x+\frac{\ell}{n}$ and $x$.  Careful evaluation of the terms results in the vanishing of the stochastic averages of (B.5) and (B.7).  The $\langle f^{a}(x+\frac{\ell}{n})\rangle_{f}$ vanishes while the other term of $\frac{df^{a}}{dx}$ combines with the other local terms.  Fortunately, the terms that will contribute $\sigma^{-\frac{3}{2}}$ cancel out and the others vanish as $\sigma\rightarrow 0$.  These results lead to equation (40).\\

Finally, terms with two derivatives are (B.2) and (B.8).  Amazingly, (B.8) gives zero because there is not enough powers of $r = (\vec{f}\cdot\vec{f})^{\frac{1}{2}}$ in the numerator to compensate the $(1+r^{2})^{4}$ in the denominator.  But the $\langle \mathbb{Y}^{ab}\rangle_{f}$ terms give
\begin{eqnarray}
\langle \mathbb{Y}^{ab}\rangle &=& \delta^{ab}\pi^{-\frac{1}{2}}(\frac{n^{2}}{\ell^{2}})\langle\sigma^{+\frac{1}{2}}\int^{\infty}_{0}\frac{e^{-\sigma r^{2}}}{(1+r^{2})^{4}} \left\{\frac{8}{3}r^{8}+8r^{6}+ \frac{8}{3}r^{4}\right\}\nonumber\\
&+& \sigma^{+\frac{3}{2}}\int^{\infty}_{0}\frac{e^{-\sigma r^{2}}}{(1+r^{2})^{4}}\left\{\frac{8}{3}r^{10}+\frac{16}{3}r^{8}+\frac{8}{3}r^{6}\right\}\rangle \label{B.15}
\end{eqnarray}
The dominant terms are the first terms of each integral and the rest all vanish as $\sigma\rightarrow 0$.  We find:
\begin{eqnarray}
\langle \mathbb{Y}^{ab}\rangle = \delta^{ab}(\frac{3}{4})(\frac{n^{2}}{\ell^{2}})\label{B.16}\
\end{eqnarray}

\section{Limited Gauge-Invariance}

Since \cite{magpantay} are unpublished, we provide some of the details in this appendix to prove the concept of limited gauge-invariance.\\

First, we prove that under condition of the positive-definiteness of $(\partial\cdot D)$, the potential given by equation (47) is invariant under infinitesimal gauge transformation.  Under gauge transformation, the LHS of equation (47) transforms as
\begin{eqnarray}
\delta A^{\prime a}_{\mu}=D^{ab}_{\mu}(A^{\prime})\sigma^{b}.\label{C.1}
\end{eqnarray}
The transformation of the RHS follows from
\begin{eqnarray}
\delta A^{a}_{\mu}=D^{ab}_{\mu}(A)\Lambda^{b},\label{C.2}
\end{eqnarray}
and using equation (49), the RHS gives
\begin{eqnarray}
RHS &=& D^{ab}_{\mu}(A)\Lambda^{b}+g\,\epsilon^{abd}(D^{de}_{\mu}(A)\Lambda^{e})_{x}\int d^{4}y\,G^{bc}(x,y;A)(\partial\cdot A^{c})_{y}\nonumber\\
&+& D^{ab}_{\mu}(A)\int d^{4}y\int d^{4}z\,G^{bd}(x,z;A)(-g)\,\epsilon^{def}(\partial\cdot D)^{fg}\Lambda^{g})_{z}G^{ec}(z,y;A)(\partial\cdot A^{c})_{y}\nonumber\\
&-& D^{ab}_{\mu}(A)\int d^{4}y\,G^{bc}(x,y;A)(\partial\cdot D^{cd}\Lambda^{d})_{y}.\label{C.3}
\end{eqnarray}
Taking the divergence of C.1 and C.3, we find that equation C.3 gives zero.  Thus, we have
\begin{eqnarray}
(\partial\cdot D^{ab}(A^{\prime}))\sigma^{b}=0.\label{C.4}
\end{eqnarray}
which is equation (50) of Section 4.  Since $\partial\cdot D(A)$ is non-singular, $\partial\cdot D(A^{\prime})$ is also non-singular because $det(\partial\cdot D)$ is invariant under infinitesimal gauge transformation.  The solution of C.4 then is $\sigma^{a}=0$, proving that $A^{\prime a}_{\mu}$ given by equation (47) is gauge-invariant.\\

Now consider the potential given by equation (53).  Following the same steps as in the previous case, we find
\begin{eqnarray}
\delta A^{\prime a}_{\mu}&=& D^{ab}_{\mu}(A^{\prime})\sigma^{b}=D^{ab}_{\mu}(A)\Lambda^{b}+g\,\epsilon^{abe}(D^{ef}_{\mu}\Lambda^{f})\int d^{4}y H^{bc}(x,y;A)[(\partial\cdot D)^{cd}(\partial\cdot A^{d})_{y}]\nonumber\\
&+& D^{ab}_{\mu}\int d^{4}y d^{4}z H^{be}(x,z,A)\delta\theta^{ef}(z)H^{fc}(z,y;A)[(\partial\cdot D)^{cd}(\partial\cdot A^{d})]_{y}\nonumber\\
&-& D^{ab}_{\mu}\int d^{4}y H^{bc}(x,y;A)\delta[(\partial\cdot D)^{cd}(\partial\cdot A^{d})]_{y}.\label{C.5}
\end{eqnarray}
In the second term in (C.5), we made use of
\begin{eqnarray}
\theta^{ab}_{x}H^{bc}(x,y;A)=\delta^{ac}\delta^{4}(x-y).\label{C.6}
\end{eqnarray}
We will also make use of the following relations in (C.5).
\begin{eqnarray}
\delta[(\partial\cdot D)^{cd}(\partial\cdot A^{d})]=\theta^{cd}\Lambda^{d},\label{C.7}
\end{eqnarray}
\begin{eqnarray}
\delta\theta^{ef}&=&-g\,\epsilon^{ekl}(D^{lm}_{\alpha}\Lambda^{m})\partial_{\alpha}(\partial\cdot D)^{kf}-g(D\cdot \partial)^{lk} \epsilon^{kfl}(\partial\cdot D)^{lm}\Lambda^{m}\nonumber\\
&-& g\epsilon^{ekh}[\partial_{\alpha}(\partial\cdot D^{hm}\Lambda^{m})]D^{kf}_{\alpha}\nonumber\\
&+& g\,\epsilon^{ekh}[\partial_{\alpha}(\partial\cdot A^{h})]\epsilon^{kfl}D^{lm}_{\alpha}\Lambda^{m}.\label{C.8}
\end{eqnarray}

Act $\partial_{\mu}$ on both sides of (C.5) followed by $(D^{a^{\prime} a}(A^{\prime})\cdot\partial)$.  When we act this last operation on the RHS of (C.5), we take note that $A^{\prime}$ must be replaced by the expression given by equation (53).  The LHS becomes $(D^{a^{\prime} a}(A^{\prime})\cdot\partial)(\partial\cdot D^{ab}(A^{\prime}))\sigma^{b}$.  The RHS is too long and rather complicated.  However, we can simplify things by neglecting terms proportional to
\begin{eqnarray*}
[\int d^{4}y H(x,y;A)[(\partial\cdot D)(\partial\cdot
A)]_{y}]^{2}\Lambda
\end{eqnarray*}
because the bracketed term is the infinitesimal gauge transformation $\rho^{a}$ (see discussions before equation (51)) that took $A^{a}_{\mu}$ to the non-linear gauge.  The neglected terms are order $\rho^{2}\Lambda$ and can thus be neglected.  With this simplification, we find
\begin{eqnarray}
(D^{a^{\prime} a}(A^{\prime})\cdot\partial)(\partial\cdot D^{ab}(A^{\prime}))\sigma^{b}=(D^{a^{\prime} a}(A)\cdot\partial)(\partial\cdot D^{ab}(A))\Lambda^{b}-g\,\epsilon^{a^{\prime} ac}D^{cd}_{\alpha}(A)\nonumber\\
\times \int d^{4}y H^{de}(x,y;A)[(\partial\cdot D)^{ef}(\partial\cdot A^{f})]_{y}\partial_{\alpha}(\partial\cdot D)^{ab}\Lambda^{b}\nonumber\\
+ g\epsilon^{abe}(D^{a^{\prime} a}\cdot\partial)(\partial\cdot D^{ef}\Lambda^{f})\int d^{4}y H^{bc}(x,y;A)[(\partial\cdot D)^{cd}(\partial\cdot A^{d})]_{y}\nonumber\\
+(D^{a^{\prime} a}\cdot\partial)(\partial\cdot D)^{ab}\int d^{4}y\int d^{4}z H^{be}(x,z;A)\delta\theta^{ef}H^{fc}(z,y;A)[(\partial\cdot D)^{cd}(\partial\cdot A^{d})]_{y}\nonumber\\
-(D^{a^{\prime} a}\cdot\partial)(\partial\cdot D^{ab})\int d^{4}y
H^{bc}(x,y;A)\theta^{cd}\Lambda^{d}\label{C.9}
\end{eqnarray}
Now, acting both sides of C.5 by $-\epsilon^{a\prime ac}[\partial_{\mu}(\partial\cdot A^{\prime c}]$, we find
\begin{eqnarray}
&-&\epsilon^{a^{\prime} ac}[\partial_{\mu}(\partial\cdot A^{\prime c})]D^{ab}_{\mu}(A^{\prime})\sigma^{b}=-\epsilon^{a^{\prime} ac}[\partial_{\mu}(\partial\cdot A^{c})]D^{ab}_{\mu}(A)\Lambda^{b}\nonumber\\
&+& \epsilon^{a^{\prime} ac}[\partial_{\mu}(\partial\cdot D)^{cd}\int d^{4}y H^{de}(x,y;A)[(\partial\cdot D)^{ef}(\partial\cdot A^{f})]_{y}]\nonumber\\
&\times& D^{ab}_{\mu}\Lambda^{b}+g\epsilon^{a^{\prime} ac}\partial_{\mu}(\partial\cdot A^{c})\epsilon^{abe}(D^{ef}_{\mu}\Lambda^{f})\int d^{4}y H^{bk}(x,y;A)[(\partial\cdot D)^{k\ell}(\partial\cdot A^{\ell})]_{y}\nonumber\\
&-& \epsilon^{a^{\prime} ac}[\partial_{\mu}(\partial\cdot A^{c})]D^{ab}_{\mu}\int d^{4}y d^{4}z H^{be}(x,z;A)\nonumber\\
&\times& \delta\theta^{ef}(z)H^{fk}(z,y;A)[(\partial\cdot D)^{k\ell}\Lambda^{\ell}]\nonumber\\
&+& \epsilon^{a\prime ac}[\partial_{\mu}(\partial\cdot
A^{c})]D^{ab}_{\mu}\int d^{4}y
H^{bd}(x,y;A)\theta^{de}\Lambda^{e}.\label{C.10}
\end{eqnarray}
Adding (C.9) and (C.10), the LHS is $\theta^{a^{\prime}
b}(A^{\prime})\sigma^{b}$.  The RHS is zero because:\\
(1)  The first term of (C.9) and the first term of (C.10) yield $\theta^{a^{\prime} }\Lambda^{b}$.\\

(2)  The fifth term of (C.9) and the fourth term of (C.10) give $-\theta^{a^{\prime} b}\Lambda^{b}$.\\

(3)  The fourth term of (C.9) and the third term of (C.10) yield\\ $\delta\theta^{a^{\prime} b}(x)\int d^{4}y H^{bc}(x,y;A)[(\partial\cdot D)^{cd}(\partial\cdot A^{d}]_{y} $.\\

Using equation C.8, this term gives
\begin{eqnarray}
&=& \left\{-g \epsilon^{a^{\prime} k \ell}(D^{\ell m}_{\alpha}\Lambda^{m})\partial_{\alpha}(\partial\cdot D)^{kb}-g(D\cdot\partial)^{a^{\prime} k}\epsilon^{kb \ell}(\partial\cdot D)^{\ell m}\Lambda^{m}\right.\nonumber\\
&-& g\epsilon^{a^{\prime} k\ell}[\partial_{\alpha}(\partial\cdot D)^{\ell m}\Lambda^{m}]D^{kb}_{\alpha}\nonumber\\
&+& \left. g\epsilon^{a^{\prime} kh}[\partial_{\alpha}(\partial\cdot A^{h}]\epsilon^{kb\ell}(D^{\ell m}_{\alpha}\Lambda^{m})\right\}\int d^{4}y H^{bc}(x,y;A)[(\partial\cdot D)^{cd}(\partial\cdot A)^{\alpha}]_{y}.\label{C.11}
\end{eqnarray}

(4)  The second term of (C.9) is cancelled by the third term of (C.11).\\

(5)  The third term of (C.9) is cancelled by the second term of (C.11).\\

(6)  The second term of (C.10) is cancelled by the first term of (C.11)\\

(7)  The third term of (C.10) is cancelled by the fourth term of (C.11).\\

Taking all these into account, we find
\begin{eqnarray}
\theta^{a^{\prime} b}(A^{\prime})\sigma^{b} = 0.\label{C.12}
\end{eqnarray}
As was shown in\cite{cal}, $\theta$ is a non-singular operator.
Thus, the only solution is $\sigma^{a} = 0$, proving that the
potential defined by equation (53) is gauge-invariant.

\end{document}